\documentclass[namedreferences]{SolarPhysics}

\usepackage[optionalrh,natbib]{spr-sola-addons}
\usepackage{graphicx}
\usepackage{color}                       
\usepackage{url}    
\usepackage{a4} 
\usepackage{solaheader}

\newcommand{\solareceiveddate}{22 Apr 2012}
\newcommand{\solaaccepteddate}{18 Sep 2012}

\renewcommand{\vec}{\mathbf}
\newcommand{\eq}[1]{(\ref{#1})}
\newcommand{\ut}[1]{\hat{\mathbf{#1}}}
\newcommand{\BE}{\begin{equation}}
\newcommand{\EE}{\end{equation}}
\newcommand{\tsum}{\sum_{m=0}^{\infty}\sum_{n=0}^{\infty}\sum_{p=0}^{\infty}}

\begin{document}
\begin{article}

\begin{opening}
\received{A} 
\accepted{1}

\title{A Magnetostatic Grad-Rubin Code for Coronal Magnetic 
       Field Extrapolations}
\author{S.A. ~\surname{Gilchrist}$^{1}$\sep
        M.S. ~\surname{Wheatland}$^{1}$}

\institute{S.A. Gilchrist \sep M.S. Wheatland \\ 
Sydney Institute for Astronomy, School of Physics, The University of Sydney, NSW 2006, Australia \\
e-mail: s.gilchrist@physics.usyd.edu.au\\}            
\begin{abstract}
The coronal magnetic field cannot be directly observed, but in principle
it can be reconstructed from the comparatively well observed photospheric
magnetic field. A popular approach uses a nonlinear force-free model. 
Non-magnetic forces at the photosphere are significant meaning the photospheric data are 
inconsistent with the force-free model, and this causes problems with 
the modeling (De Rosa {\it et al.}, {\it Astrophys. J. } {\bf 696}, 1780, 2009).
In this paper we present a numerical implementation of the Grad-Rubin method 
for reconstructing the coronal magnetic field using
a magnetostatic model. This model includes a pressure force and
a non-zero magnetic Lorentz force. We demonstrate our 
implementation on a simple analytic test case and obtain the 
speed and numerical error scaling as a function of the grid size. 
\end{abstract}
\end{opening}


%
%

\section{Introduction}
\label{introduction}

The solar magnetic field is observed primarily through
its effect on the polarization of particular Zeeman sensitive spectral 
lines \citep{2004ASSL..307.....L}. Observations of the Fe {\sc i}  
multiplet provide two-dimensional maps of the vector 
magnetic field (vector magnetograms) close to the height of the photosphere. 
Unlike the photosphere, the corona lacks suitable magnetic lines 
and the coronal magnetic field cannot be determined   
except under exceptional circumstances ({\it e.g.} \citealt{1964ApJ...140..817H}),
although new methods based on radio and infrared observations are 
presently being developed ({\it e.g.} \citealt{1997SoPh..174...31W,2004ApJ...613L.177L}). 
The inability to observe the coronal field presents a 
barrier to understanding important coronal magnetic phenomena, including
solar flares.

In principle, the coronal magnetic field can be 
reconstructed from photospheric magnetograms
by using a time-independent magneto-hydrodynamic model of the corona. 
The model requires solution of equations for the coronal magnetic field, $\vec B$,
subject to boundary conditions derived from vector 
magnetogram data. The field obtained by solving the model is
a proxy for observational data. How accurately the model field reflects the 
true coronal field depends on the quality of the magnetogram
data, and the accuracy of the assumptions in the model.  

Non-magnetic forces in the corona are generally
negligible \citep{1995ApJ...439..474M,2001SoPh..203...71G} 
and this motivates coronal magnetic field reconstructions 
based on a force-free model, {\it i.e.} one in which only the 
magnetic (Lorentz) force is considered.
In the force-free model the local current density is proportional to 
the magnetic field \citep{1984smh..book.....P}, with the proportionality 
factor $\alpha$ in general a function of position. The special case where
$\alpha$ is independent of position is called a linear force-free
field. The linear force-free model can be solved analytically and has been 
extensively studied ({\it e.g.} \citealt{1972SoPh...25..127N,1978SoPh...56...55B,1978SoPh...58..215S,1981A&A...100..197A}). 
However the linear force-free model is of limited use 
as a model of the coronal magnetic
field because of specific unphysical features ({\it e.g.}
a linear force-free field in an unbounded domain has
infinite magnetic energy \citep{1981A&A...100..197A}). 

The problems with the linear force-free model
spur interest in the nonlinear force-free model, where
$\alpha$ varies with position. The formal boundary value problem 
for force-free modeling can be stated in different ways depending on 
the choice of boundary conditions on the photosphere (different 
possible choices are discussed by \citet{gr} and \citet{2007GApFD.101..249A}). 
Some definitions use all three components of $\vec B$ over the entire boundary,
while others prescribe the normal component of the magnetic field 
$\vec B \cdot \ut{n}$ and the distribution of $\alpha$ over a single
polarity of $\vec B \cdot \ut{n}$, {\it i.e.} where $\vec B \cdot \ut{n}<0$ 
or where $\vec B \cdot \ut{n}>0$. The latter has been shown to be 
a well-posed\footnote{A boundary value problem is said to be well-posed
if it has a unique solution which depends continuously on the boundary
conditions.} formulation of the problem (in certain domains) for
sufficiently small values of $\alpha$, while the former is an over 
specification of the problem \citep{bineau,2000ZaMP...51..942B,2000CMaPh.211..111K}. 
In both cases the boundary value problem is nonlinear
and in general requires a numerical treatment. A number of different 
numerical solution methods have been developed for different statements
of the nonlinear force-free boundary value problem 
(see reviews by \citet{1989SSRv...51...11S}, or more recently 
\citet{2008JGRA..11303S02W}). \citet{2007GApFD.101..249A}
review the different methods and the choice of boundary conditions used
in each case.

Although appealing in its simplicity, the application of the (nonlinear)
force-free model to magnetogram data has proved 
problematic ({\it e.g.} \citealt{2009ApJ...696.1780D,2008ApJ...675.1637S}). 
The numerical solution methods are generally iterative, 
and when applied to solar data the iterations may fail to converge \citep{2009ApJ...696.1780D}. 
The problem is particular pronounced when large electric 
currents are present in the data. (It should be noted that convergence
problems were not reported by \citet{2009ApJ...696.1780D} for two of 
the methods which are based on a well-posed formulation of the 
boundary value problem.) 

A possible reason for the problems is that the 
data departs significantly from a force-free state
\citep{2001SoPh..203...71G,1995ApJ...439..474M}. This is expected
to affect different solution methods in different ways. For methods 
using the well-posed formalism, two solutions are obtained for a 
single magnetogram corresponding to the two choices of polarity.
In practice, these two solutions are qualitatively and quantitatively
different due to the forced nature of the boundary data. 
For methods which over specify the problem the iterative methods
can never converge to a force-free state if the boundary data are
not consistent with the force-free model. In practice, this leads 
to numerical solutions with residual forces and nonzero divergence.
A proposed solution to this problem is `preprocessing' of the 
data \citep{2006SoPh..233..215W,2007A&A...476..349F}. 
With this technique the data are modified to be consistent with
small net forces and torques in the overlying volume while maintaining
minimal departure from the original data. However, the preprocessed data
are not necessarily consistent with the 
force-free model \citep{2009ApJ...696.1780D} and
mixed results have been obtained in the modelling: in some cases the problems
were reduced \citep{2008ApJ...675.1637S} and in other cases there was no 
significant improvement \citep{2009ApJ...696.1780D}. 

Another approach is to develop static models of the coronal magnetic
field which incorporate non-magnetic forces, {\it i.e.} magneto-hydrostatic
models, which incorporate gravity and gas pressure forces. The special
case of a model with only magnetic and pressure forces may be called 
a magnetostatic model. This model requires the prescription of pressure 
as a boundary condition. Presently, the photospheric pressure distribution
is difficult to obtain observationally compared to  
the magnetic field. This is a limitation of magnetostatic (and
magneto-hydrodynamic) models in application to solar data.

A number of methods for solving the magneto-hydrostatic and magnetostatic
equations have been developed for modeling the coronal
magnetic field.  The optimization procedure developed for the nonlinear force-free 
model has been extended to magneto-hydrostatic and magnetostatic
models in Cartesian geometry \citep{2003SoPh..214..287W,2006A&A...457.1053W},
and spherical geometry \citep{2007A&A...475..701W}. \citet{amari2009} 
presented a finite element implementation of the Grad-Rubin method \citep{gr} for 
solving a magnetostatic model which is applicable in arbitrary 
geometry. In addition semi-analytic approaches have been
developed but these provide only restricted, and not general, 
solutions ({\it e.g.} \citealt{2001SoPh..198..279R}).

In this paper we present a numerical code to 
solve a general magnetostatic
model of the coronal magnetic field of an active region. 
Our code is an implementation of the Grad-Rubin method \citep{gr} to 
the magnetostatic equations in the half-space $z\ge 0$. The Grad-Rubin 
method has previously been used to solve the nonlinear force-free
equations in the half-space for coronal reconstructions 
({\it e.g.} \citealt{2006SoPh..238...29W,1999A&A...350.1051A}), 
and to solve the magnetostatic equations
in toroidal geometry \citep{1986CoPhC..43..157R} and in arbitrary geometry
\citep{amari2009}. The Grad-Rubin method has the
advantage of, in principle, solving the magnetostatic equations
to the limited imposed by numerical accuracy ({\it e.g.} truncation error
due to the finite grid size), assuming the Grad-Rubin iteration
procedure converges. We apply the code to a simple analytic test case as a 
demonstration, and to investigate the speed and numerical error
scaling of the implementation. The resulting code and method does not yet
provide a practical tool for coronal field modeling from solar data, 
because of the neglect of the gravity force, but it is a significant step
in this direction.    

This paper is structured as follows. Section \ref{theory} presents
the magnetostatic equations, and the boundary value problem to
be solved. Section \ref{code} is a brief summary of the
Grad-Rubin method, and describes our specific implementation of the 
method in code. Section \ref{test_case} presents the analytic test case 
used. Section \ref{results} presents the results, and Section 
\ref{discussion} contains discussion of the results and the conclusion.

%
%

\section{Magnetostatic Equations and Boundary Value Problem}
\label{theory}

\subsection{The Model}
\label{The_Model}

The magneto-hydrostatic equations with the gravity force neglected
are 

\begin{equation}
 \nabla \times \vec B  = \mu_0 \vec J, 
 \label{ampere}
\end{equation}

\begin{equation}
 \nabla \cdot \vec B = 0,
 \label{gauss}
\end{equation}
and
\begin{equation}
 \vec J \times \vec B - \nabla p = 0
 \label{force}
\end{equation}
\citep{1984smh..book.....P}. Here $p$ is the gas pressure, $\vec J$ is 
the electric current density, and $\vec B$ is the magnetic field vector.

\subsection{The Boundary Value Problem}
\label{BVP}

For a localized active region the curvature of the photosphere
is small, so we solve Equations \eq{ampere}-\eq{force} in the 
half-space $z \ge 0$, with the $z=0$ plane representing
the photosphere. The appropriate boundary conditions prescribed on the 
$z=0$ plane are $B_z$, together with $p$ and $J_z$ prescribed over a 
single polarity of the magnetic field, {\it i.e.} $p$ and $J_z$ are prescribed 
only at points with $B_z>0$ or at points where $B_z<0$ \citep{gr}. 
We denote the boundary values of the magnetic field, pressure and
current density by $B_{\rm obs}$, $p_{\rm obs}$ and 
$J_{\rm obs}$ respectively. These boundary conditions are believed 
to be the correct physical boundary conditions for the magnetostatic
model. 

Equations \eq{ampere}-\eq{force} are solved numerically in the 
finite volume
\BE
  \Omega_b =  \{(x,y,z) |\, 0 \le x \le L_x, 0 \le y \le L_y, 0 \le z \le L_z\}.
\EE
In addition to the $z=0$ plane, $\Omega_b$ has five additional plane boundaries
on which boundary conditions are required. Magnetogram data 
only provide boundary conditions on the $z=0$ plane and 
reasonable assumptions must be made for the remaining five.
This problem is faced by all reconstruction codes regardless of the
particular model or method used, and models including more physics 
typically require more boundary conditions at each boundary. 

We choose boundary conditions on the magnetic field such that 
all field lines are connected to the lower boundary at two points 
({\it i.e.} there are no open field lines). The need for this 
 is discussed in Section \ref{trace}. 
We achieve this in practice
by imposing either (i) closed boundary conditions on the top and side 
boundaries
\BE
  \vec B \cdot \hat{\mathbf n} =0, 
\EE
where $\hat{\mathbf n}$ denotes the unit normal vector to each 
boundary, or (ii) a closed top boundary condition
\BE
  \vec B \cdot \hat{\mathbf z} =0,
\EE 
together with periodic boundary conditions on the side boundaries. 
For the periodic case any field line which leaves the computational
volume by a side boundary re-enters on the opposite side, and therefore 
eventually connects to the lower boundary. These boundary conditions specify that 
magnetic flux only enters and leaves through the lower boundary, which requires 
that the lower boundary is flux balanced.

%
%

\section{Numerical Implementation of the Grad-Rubin Method}
\label{code}

In this section we outline our implementation of the Grad-Rubin 
method in code. The approach is similar to that for the 
force-free code described in \citet{2007SoPh..245..251W}. 

\subsection{The Grad-Rubin Method}
\label{gr_iteration}

The Grad-Rubin method (also called the current-field iteration method) 
is an iterative scheme for solving the magnetostatic 
equations \citep{gr}. In this method the nonlinear 
equations (Equations (\ref{ampere})-(\ref{force})) are replaced with a 
set of linear equations which are solved at each iteration. 

Here we briefly outline a single Grad-Rubin iteration (a more detailed
description is given by \citet{gr}). We denote a quantity after
$k$ Grad-Rubin iterations using a superscript, so for example 
$\vec B^{(k)}$ denotes the magnetic field in our computational volume
after $k$ iterations starting from an initial magnetic 
field $\vec B^{(0)}$. In practice the iteration is initiated 
with a potential field in the volume calculated from 
$B_{\rm obs}$, which we denote as $\vec B^{(0)}= \vec B_{0}$. 
In the following we assume $\vec B^{(k)}$ is known from a previous 
iteration, or is the initial potential field. A single iteration
consists of the following steps.

\begin{enumerate}
 \item Calculate a new pressure $p^{(k+1)}$ in the volume by solving
   \BE
     \vec B^{(k)} \cdot \nabla p^{(k+1)}   = 0, \label{gr_s1}  
   \EE  
   with boundary conditions 
   \BE
     p^{(k+1)}|_{z=0} = p_{\rm obs} \label{gr_bc1} 
   \EE
    prescribed over one polarity of $B_{\rm obs}$.
  \item Calculate the component of the current density perpendicular 
   to the magnetic field in the volume using 
   \BE
     \vec J_{\perp}^{(k+1)}  = \nabla p^{(k+1)} \times \vec B^{(k)}/|\vec B^{(k)}|^2.\label{gr_s2} 
   \EE 
   \item Calculate the component $\vec J_{\parallel}^{(k+1)}$ 
   of the current density parallel to the magnetic field in the
   volume. The parallel component can be written as 
   \BE
     \vec J_{\parallel}^{(k+1)} = \sigma^{(k+1)} \vec B^{(k)}/\mu_0,
     \label{jpar}
   \EE 
   where $\sigma^{(k+1)}$ is a scalar function of position. 
   The parameter $\sigma^{(k+1)}$ is calculated in the volume by solving \citep{gr}
   \BE
     \nabla \sigma^{(k+1)} \cdot \vec B^{(k)} = -\mu_0 \nabla \cdot \vec J^{(k+1)}_{\perp}, 
     \label{gr_s3}
   \EE
   with boundary conditions 
   \BE
     \sigma|_{z=0} = \sigma_{\rm obs}^{(k+1)}
   \EE
   where
   \BE  
     \sigma_{\rm obs}^{(k+1)} = \left. \frac{ \mu_0 
     (\vec J_{\perp}^{(k+1)} \cdot \hat{\mathbf z} - J_{\rm obs})}
     {B_{\rm obs}} \right |_{z=0}. \label{alpha_obs}
   \EE
   The boundary conditions $\sigma_{\rm obs}^{(k+1)}$ are prescribed
   over a single polarity of $B_{\rm obs}$ and are calculated using 
   Equation \eq{alpha_obs} at each iteration. The form of Equation \eq{alpha_obs} is 
   such that $\vec J^{(k+1)} \cdot \ut{z} = J_{\rm obs}$ on the lower
   boundary (over the polarity of $B_{\rm obs}$ chosen for the boundary 
   conditions).
  
   Equations \eq{gr_s2}-\eq{jpar} define the total current density in 
   the volume:
   \BE
     \vec J^{(k+1)} = \vec J^{(k+1)}_{\perp} + \sigma^{(k+1)} \vec B^{(k)}/\mu_0.
   \EE   

   \item Calculate the new magnetic field in the volume by solving
   Ampere's law 
   \BE
     \nabla \times \vec B^{(k+1)} = \mu_0 \vec J^{(k+1)},
     \label{gr_ampere}
   \EE
   where the boundary conditions on $\vec B^{(k+1)}$ are those in Section \ref{theory}. 
\end{enumerate}

\subsection{Overview of the Code}

The code solves the magnetostatic equations in the 
finite Cartesian domain $\Omega_{\rm b}$ using the Grad-Rubin method. 
The numerical grid is 
uniformly spaced with $N$ grid points along each 
dimension, so there are $N^3$ grid points in total, and in the following
we write $L=L_x=L_y=L_z$ for simplicity. The grid spacing 
is $L/(N-1)$, and the coordinates of the grid
points are $(x_i,y_j,z_k) =  (i,j,k)L/(N-1)$, where 
$0 \le i,j,k \le N-1$. 

In the following we describe the implementation of the Grad-Rubin
steps identified in Section \ref{gr_iteration} in detail, identifying
some of the numerical methods used. The code is an implementation in 
FORTRAN90 using double precision floating point numbers \citep{1992nrfa.book.....P}. 
The code is parallelized for shared memory multiprocessors using OpenMP \citep{openmp}.

\subsubsection{Step 1: Update of Pressure in the Volume}
\label{trace}

Equation \eq{gr_s1} is solved by a field line tracing (or characteristic)
method. The same method is used in some existing force-free Grad-Rubin
method implementations for solving 
$\vec B \cdot \nabla  \alpha =0$ ({\it e.g.} \citealt{1999A&A...350.1051A,2006SoPh..238...29W,2007SoPh..245..251W}),
where $\alpha$ is the force-free parameter defined by  
$\nabla \times \vec B = \alpha \vec B$. Equation \eq{gr_s1} has also 
been solved using a finite element method \citep{2006A&A...446..691A}. In 
the force-free case the tracing is used to update $\alpha$ in the 
volume, and here it is used to update the pressure. 

The procedure is as follows. For each grid point $(x_i,y_j,z_k)$ the field line threading 
$(x_i,y_j,z_k)$ is traced to the point $(x_0,y_0)$ where it crosses the 
lower boundary with the appropriate polarity for the boundary conditions. 
The pressure at the grid point is then assigned to be equal to the 
boundary value:  
\BE
  p(x_i,y_j,z_k) = p_{\rm obs}(x_0,y_0).  
\EE
This procedure solves Equation \eq{gr_s1} because the pressure 
is constant along a magnetic field line. The tracing is performed in either 
the forward or backward direction along the field line, with 
the direction chosen such that the point $(x_0,y_0)$ has
the appropriate polarity for $B_{\rm obs}$. Fourth order Runge-Kutta
is used for the numerical tracing \citep{1992nrfa.book.....P}. 
Because the path of the field line is not confined to grid points, 
trilinear interpolation is used to estimate $\vec B$ between 
grid points \citep{1992nrfa.book.....P}. Similarly, bilinear 
interpolation is used to compute $p_{\rm obs}$ at the boundary 
point $(x_0,y_0)$, which also may not coincide with a grid point.

Open field lines (see Section \ref{BVP}) would present a problem for 
this method. Since $p_{\rm obs}$ is only prescribed over a 
single polarity, it would be impossible to assign pressure to 
points threaded by field lines which connect to the lower boundary at
only one polarity, opposite to that for which $p_{\rm obs}$ is 
prescribed. We avoid this problem by preventing open field lines 
through the choice of boundary conditions on $\vec B$ 
explained in Section \ref{BVP}. This eliminates
the problem but introduces artificial boundary conditions on the 
top and side boundaries. However, the region of interest can 
be isolated from the effects of the boundaries by using a large
domain. 

\subsubsection{Steps 2 and 3: Update of the Current Density in the Volume}
\label{steps_23}

Given the updated values $p^{(k+1)}$ for the pressure in the
volume, Equation \eq{gr_s2} may be directly evaluated at every gridpoint
to give the perpendicular current $\vec J_{\perp}$ in the volume.
The derivatives in the gradient on the right hand side are evaluated
numerically using a centered difference approximation \citep{1992nrfa.book.....P}.

A particular problem is encountered with the evaluation of the 
perpendicular current $\vec J_{\perp}$ in the volume at
each iteration (step 2 in the enumeration of the procedure in
Section \ref{gr_iteration}), for the test cases considered here. The 
perpendicular current is calculated using Equation \eq{gr_s2}:
\BE
  \vec J_{\perp}^{(k+1)}  = \nabla p^{(k+1)} \times 
  \vec B^{(k)}/|\vec B^{(k)}|^2. 
\EE 
For the analytic solutions we use in Section \ref{results} there are locations in the volume 
where $\vec B =0$ and $\vec J_{\perp}$ is finite. The perpendicular 
current is not correctly evaluated numerically at these points. To 
prevent this we choose grid sizes for our problems such that the 
points with $\vec B=0$ fall between grid points. This removes 
the problem, which is due to the artificial nature of the
test cases.

The value of $\sigma^{(k+1)}$ at each gridpoint is then obtained by 
solving Equation \eq{gr_s3}. This equation may be 
integrated along a field line to give the formal solution
\BE
 \sigma^{(k+1)}(x_i,y_j,z_k) =  \sigma_{\rm obs}(x_0,y_0) - \gamma \int_0^{s_0} \nabla \cdot 
 \vec J_{\perp}^{(k+1)}(\vec x(s))/|\vec B^{(k)}(\vec x(s))|ds, 
\EE
where
\begin{equation}
\gamma = \left \{ \begin{array}{rl} 
& +1 \;  \mathrm{if}\; J_{\rm obs}\; \mathrm{is\; prescribed\; over}\; B_z>0 \\
& -1 \; {\rm if}\; J_{\rm obs}\; {\rm is \; prescribed \; over}\; B_z<0, \\
 \end{array} \right. 
\label{alpha_sol}
\end{equation}
where $\vec x(s)$ is the path of the field line (the parameter $s$ is
the arc length along the field line), and where $\vec x(s_0)$ is the
point $(x_0,y_0,0)$ at which the boundary conditions on $\sigma^{(k+1)}$
are imposed. Trilinear interpolation is used to assign values in the 
argument of the integral along a field line, and the integral is evaluated 
using the trapezoidal rule \citep{1992nrfa.book.....P}. 

The field line tracing needed for steps one and two (updating pressure
and $\sigma$) is the computationally slow part of this implementation
of the Grad-Rubin method. The number of operations for the field line tracing scales as $N^4$
for a grid with $N^3$ points \citep{2006SoPh..238...29W}. In the
following we will write `$\sim N^4$' to denote such a scaling.
The code parallelizes the process using OpenMP, with the workload 
divided such that different code threads trace different field lines. 

It is important to understand the accuracy of the numerical
solutions for $p^{(k)}$ and $\sigma^{(k)}$. For the field line tracing solution 
we can infer an approximate scaling for the numerical error as follows. 
Trilinear interpolation has truncation error $\sim 1/N^2$ \citep{zik2010}, and on average a field line requires 
$\sim N$ Runge-Kutta steps to reach the lower boundary. Therefore the 
total numerical error introduced by tracing a field line to the lower
boundary has scaling $\sim N \times 1/N^2 = 1/N$). 
We expect this error to be the dominant error in the calculation, and 
so the error scaling for the whole computation is $\sim 1/N$. 
This error scaling has been confirmed for the force-free 
case \citep{2006SoPh..238...29W}.

\subsubsection{Step 4: Update of the Magnetic Field in the Volume}
\label{mag_field}

The magnetic field may be expressed as $\vec B^{(k)}=\vec B_0 + \vec B_c^{(k)}$,
where $\vec B_0$ is the potential field satisfying $\nabla \times \vec B_0=0$
together with the boundary condition

\BE
 \vec B_0 \cdot \hat{\mathbf z}|_{z=0} = B_{\rm obs},
 \label{bc1}
\EE
and where $\vec B_c^{(k)}$ is a current carrying field satisfying 
$\nabla \times \vec B_c^{(k)} = \mu_0 \vec J^{(k)}$ together with the 
homogeneous boundary condition

\BE
 \vec B_c^{(k)} \cdot \hat{\mathbf z}|_{z=0} = 0.
\EE
The fields $\vec B_c^{(k)}$ and $\vec B_0$ 
have the same boundary conditions on the top and side boundaries
(either closed boundary conditions on all other boundaries, or closed top boundary
conditions and periodic side boundary conditions, as discussed in 
Section \ref{BVP}).

For the potential field a scalar potential $\phi$ can be introduced 
defined by 
\BE
  \nabla \phi = - \vec B_0,
\EE
and the problem reduces to solving Laplace's equation \citep{1998clel.book.....J}:
\BE
  \nabla ^2 \phi = 0.
\EE
Laplace's equation has well-known Fourier solutions in Cartesian 
geometry \citep{1953mtp..book.....M}, and we 
use a two-dimensional Fourier series 
solution with the expansion performed in the $x$ and $y$ directions.
For periodic boundary conditions the components of our potential 
field are\footnote{We note that this solution is a special case 
of the linear force-free solution due to 
\citet{1978SoPh...56...55B}, and is obtained by setting $\alpha=0$ in that
solution.}

\begin{eqnarray}
B_{0x}(x,y,z) &=& \sum_{m=0}^{\infty}\sum_{n=0}^{\infty} c_{mn} i k_m
                  \cosh(k[z-L]){\rm e}^{i(xk_m+yk_n)},\label{b01} \\
B_{0y}(x,y,z) &=& \sum_{m=0}^{\infty} \sum_{n=0}^{\infty} c_{mn} 
                  i k_n \cosh(k[z-L]){\rm e}^{i(xk_m+yk_n)}, \label{b02}
\end{eqnarray}
and
\BE
B_{0z}(x,y,z) = \sum_{m=0}^{\infty} \sum_{n=0}^{\infty} c_{mn} k
                  \sinh(k[z-L]){\rm e}^{i(xk_m+yk_n)}\label{b03}. 
\EE
where $k_m = 2\pi m/L$, $k_n = 2\pi n/L$ and $k^2 = k_m^2+k_n^2$. 
The Fourier series coefficients are derived from the boundary 
conditions:
\BE
 c_{mn} = -\frac{1}{L^2} \int_0^{L} 
 \int_0^{L} dxdy B_{\rm obs}(x,y) {\rm e}^{-i(xk_m+yk_n)}/\sinh(kL).
\EE
As explained in Section \ref{BVP}, we also use a solution with 
closed boundaries, and the expression for the components of this
field are similar (see Appendix A).
Equations \eq{b01}-\eq{b02} can be computed using Fast Fourier
transforms, in which case $\sim N^3 \log(N)$ operations are required to 
evaluate the potential field for a grid of $N^3$ grid 
points \citep{1992nrfa.book.....P}. This makes step 4 relatively fast,
computationally.

For the non-potential component $\vec B_c$ we use a three-dimensional
Fourier series solution to the vector Poisson equation, working with 
a vector potential $\vec A_c$ such that 
$\vec B_c =\nabla \times \vec A_c$ \citep{1953mtp..book.....M}. 
The components of the field are 

\begin{eqnarray}
B_{cx}(x,y,z) &=& \tsum 
                  \left [k_nia^{(3)}_{mnp}-k_pa^{(2)}_{mnp} \right ]
                  \cos(k_p z){\rm e}^{i(k_m x+k_n y)}/k^2,\label{np_b1} \\
B_{cy}(x,y,z) &=& \tsum 
                  \left [k_pa^{(1)}_{mnp}-k_mia^{(3)}_{mnp} \right ]
                  \cos(k_p z){\rm e}^{i(k_m x+k_n y)}/k^2, \\
\end{eqnarray}
and
\BE
B_{cz}(x,y,z) =   \tsum 
                  i\left [k_m a^{(2)}_{mnp}-k_n a^{(1)}_{mnp}\right ]
                  \sin(k_p z){\rm e}^{i(k_m x+k_n y)}/k^2. \label{np_b3}   
\EE
Here $k_m = 2\pi m/L$, $k_n = 2\pi n/L$, and $k_p = \pi p/L$.
The coefficients $a^{(i)}_{mnp}$, with $i=1,2,3$ are given by:
\begin{eqnarray}
  a^{(1)}_{mnp} & =& \frac{2\mu_0}{L^3} 
    \int_0^{L}\int_0^{L}\int_0^{L} J_x(x,y,z) 
    {\rm e}^{-i(k_m x + k_n y)}\sin(k_p z)dxdydz,\label{np_a1} \\
  a^{(2)}_{mnp} & =& \frac{2\mu_0}{L^3} 
    \int_0^{L}\int_0^{L}\int_0^{L} J_y(x,y,z) 
    {\rm e}^{-i(k_m x + k_n y)}\sin(k_p z)dxdydz, \label{np_a2}
\end{eqnarray}
and
\BE
  a^{(3)}_{mnp} = \frac{2\mu_0}{L^3} 
    \int_0^{L}\int_0^{L}\int_0^{L} J_z(x,y,z)
    {\rm e}^{-i(k_m x + k_n y)}\cos(k_p z)dxdydz. \label{np_a3}
\EE
The coefficients given by Equations \eq{np_a1}-\eq{np_a3}, and the solution
given by Equations \eq{np_b1}-\eq{np_b3}, are computed in $\sim N^3\log(N)$
operations using a combination of fast Fourier, fast sine, 
and fast cosine transforms \citep{1996tah..book.....P}. 
The corresponding solution for $\vec B_c$ with closed 
side boundaries is given in Appendix B, which also provides more detail
on how these solutions are derived.

%
%

\section{Analytic Test Case}
\label{test_case}

To test the code we use a simple analytic solution
to the magnetostatic equations 
\citep{1998PhST...74...77W,2003SoPh..214..287W} which is a generalization 
of a force-free field in \citet{1994ppit.book.....S}, and which may 
be derived using the generating function method. The solution describes 
a sheared magnetic arcade with translational 
symmetry in the $y$ direction and periodicity in the $x$ direction. 

For the problem at hand we modify the sheared arcade solution 
by imposing a closed top boundary condition (to match the boundary 
condition required by our Grad-Rubin method):  
\BE
  B_z(x,y,L)=0. 
\EE
The components of the resulting magnetostatic field are    
\begin{eqnarray}
 B_x(x,z) &=& \psi_0 l \sin(kx)\cosh[l(L-z)], \\ 
 B_y(x,z) &=& \psi_0 \lambda \sqrt{1-a_0}  \sin(kx)\sinh[l(L-z)],
\end{eqnarray}
and
\BE
  B_z(x,z) = \psi_0 k \cos(kx)\sinh[l(L-z)], \label{an_bz}
\EE
and the pressure is given by
\BE
 p(x,z) = \psi_0^2 \left (\frac{ a_0 \lambda}{2\mu_0} \right ) 
 \sin^2(kx)\sinh^2[l(z-L)],
\EE
where $k$, $\lambda$, $\psi_0$ and $a_0$ are free parameters subject to
$0 \le a_0 \le 1$ and $l = \sqrt{k^2-\lambda^2}$. 

The parameter $\lambda$ determines the currents in the volume,
with $\lambda=0$ giving a potential field. The parameter
$a_0$ sets the pressure in the volume, and the special case $a_0=0$  
is a linear force-free field with force-free 
parameter $\alpha=\lambda$. The parameter $a_0$ may be used to 
enforce closed side boundary conditions in $y$ with the choice $a_0=1$. 
The parameter $k$ determines the period of the solution in the
$x$ direction, and the constant $\psi_0$, which determines the magnitude
of the field, is chosen to be
\BE
 \psi_0 = \frac{1}{\sinh(l L)k},
 \label{norm1}
\EE
to specify $\mbox{max}(|B_{\rm obs}|)=1$ on the lower boundary.

The boundary conditions in the $z=0$ plane with the choice 
of $\psi_0$ given by Equation \eq{norm1} are

\begin{eqnarray}
  B_{\rm obs} &=& \cos(kx), \label{bobs}\\
  p_{\rm obs} &=& \frac{a_0\lambda}{2\mu_0 k^2} \sin^2{(kx)} \label{pobs},
\end{eqnarray}
and 
\BE
  J_{\rm obs} = \mu_0\lambda \sqrt{1-a_0} \cos{(kx)} \label{jobs}.
\EE

\begin{figure}[!h]
\centerline{\includegraphics[scale =0.77]{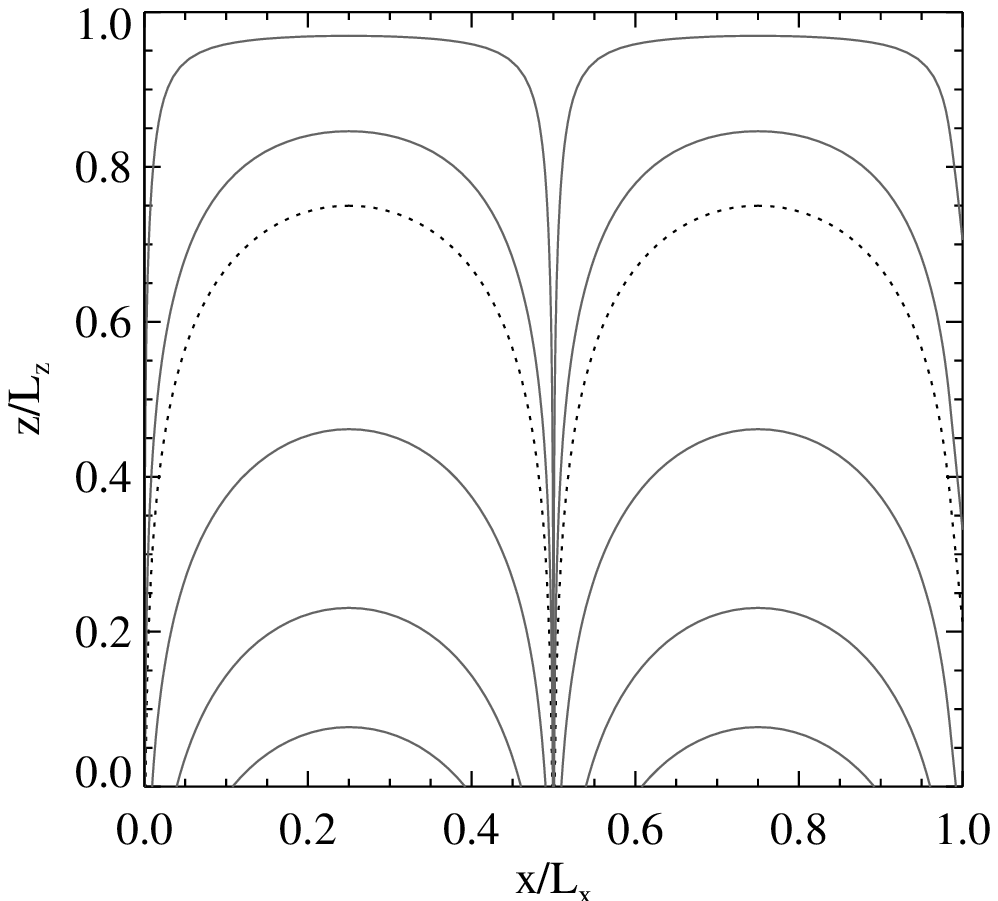},\includegraphics[scale =0.77]{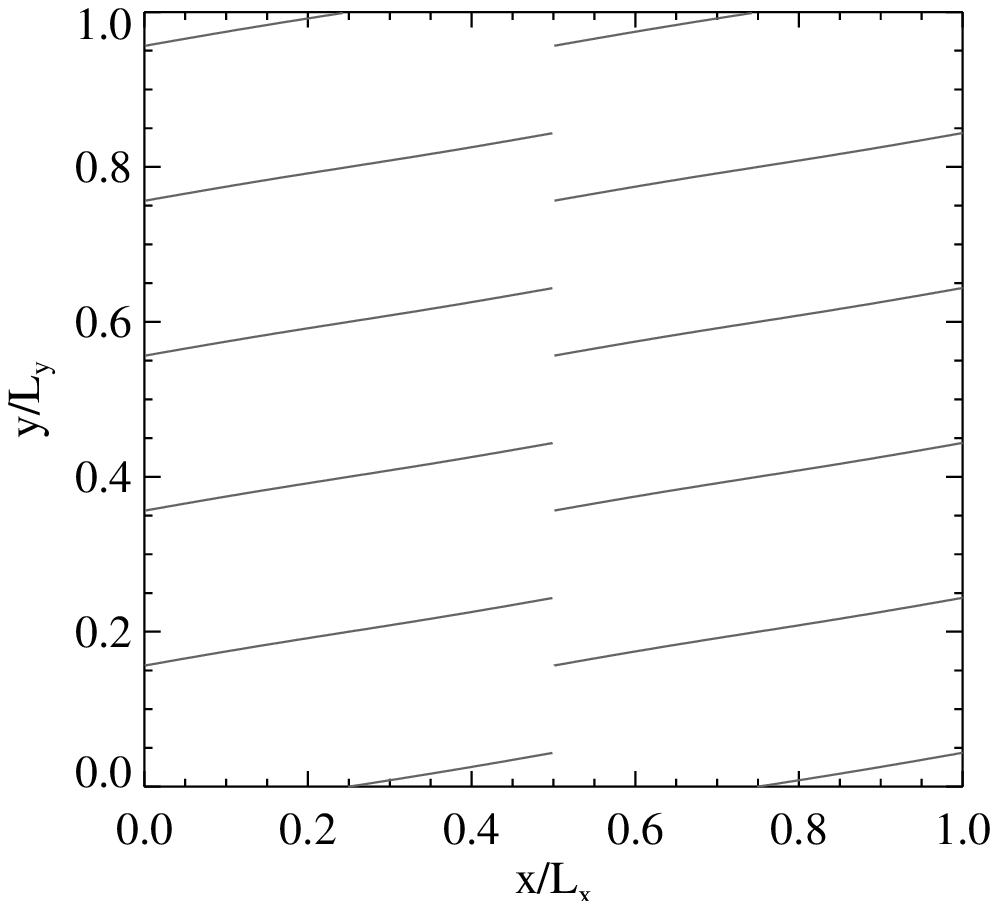}}
\caption{Field lines for the analytic magnetostatic field test case (see 
Section \ref{test_case}) with parameters $k=2\pi/L$, $a_0=0.5$, and 
$\lambda= \pi/(2L)$. In the left panel the point of view is 
along the $y$ axis. The right panel shows a top-down view of 
the field lines indicated by the dashed curves in the left panel. These
field lines are sheared with respect to the $x$ axis, at an angle of
$\approx 10^{\circ}$.}
\label{f1}
\end{figure}

A schematic diagram of the field lines of the solution is shown in Figure \ref{f1},
for the choices $k=2\pi/L$, $a_0=0.5$, and $\lambda = \pi/(2L)$.
The view in the left panel of the Figure is along the $y$ axis. 
This perspective shows the arcade-like field line structure.
The right panel of Figure \ref{f1} shows a top-down view of the particular
field lines shown as dashed curveswh in the left panel. The top
of these field lines is at $z= 3L/4$. This 
perspective shows that the arcade is sheared. For the given solution 
the shear angle is $\approx 10^{\circ}$ for the field lines with the 
height shown. 

%
%
\section{Results}
\label{results}

In this section we apply the code to the test case in 
Section \ref{test_case} for two different choices of parameters. The first
choice is for a calculation with periodic side boundary conditions, and the
second is for a calculation with closed side boundary conditions. 
For both cases the tests are performed several
times on grids of varying size. The convergence of 
the Grad-Rubin iteration is demonstrated, and the speed 
of the code as a function of problem size (introduced in Section \ref{steps_23}) 
is confirmed. We also examine the accuracy of the numerical solution by 
comparison with the analytic solution, and investigate the accuracy
as a function of grid size, for comparison with the
estimate of the scaling of the accuracy given in Section \ref{steps_23}.  

\subsection{Test Case with Periodic Side Boundaries}
\label{TC1}

For the first test we use the parameters
$k=2\pi(1-1/N)/L$, $\lambda=\pi/(2L)$ and $a_0=0.5$. The parameter $k$ is 
chosen to vary with $N$ so that the periodicity of the 
solution matches the periodicity of the discrete 
Fourier transform \citep{1992nrfa.book.....P}. 
A side effect of this is that the test case differs between grid sizes. In particular 
the boundary conditions, which depend on 
$k$ through Equations \eq{bobs}-\eq{jobs}, are  different. 
However, for large $N$ the difference is small. We apply the code 
to the test case with three different grid sizes ($N=51,101,151$), 
and in each case we perform 30 Grad-Rubin iterations 
starting from a potential field.

\begin{figure}[!h] 
\centerline{\includegraphics[scale =0.75]{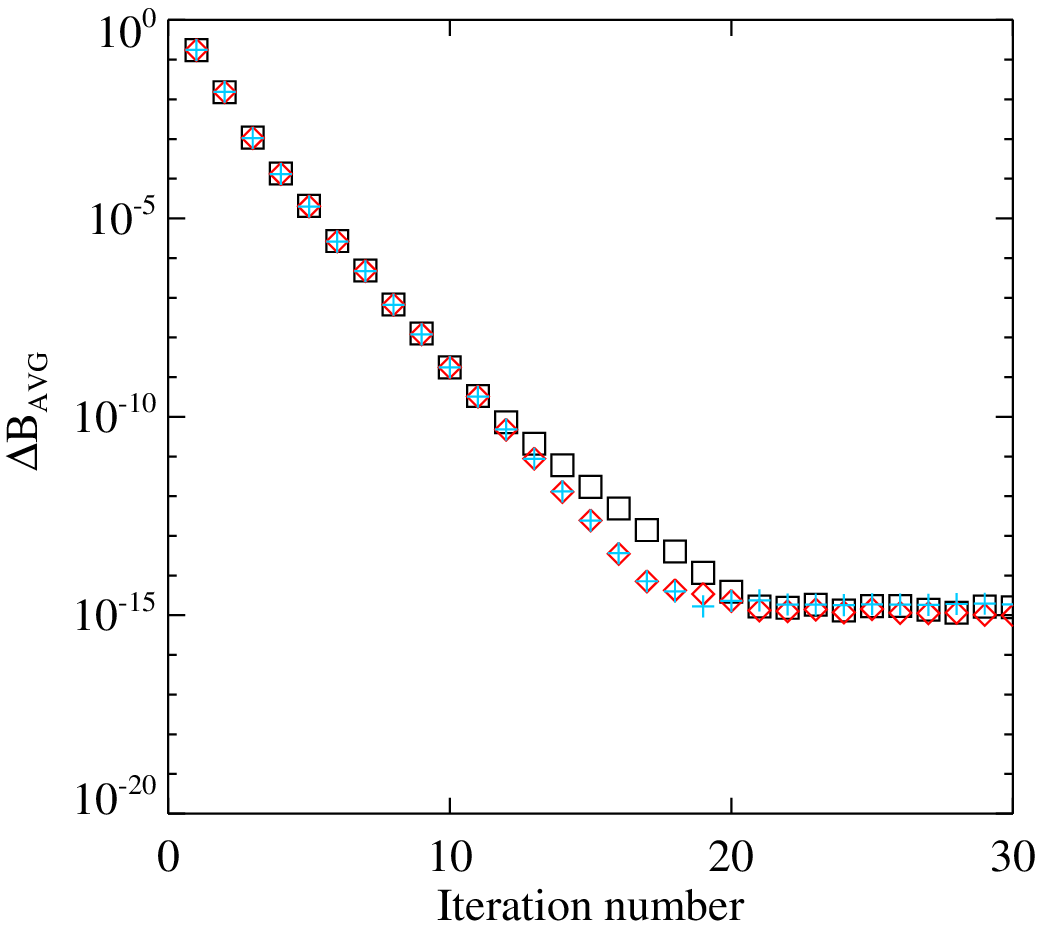} \includegraphics[scale =0.75]{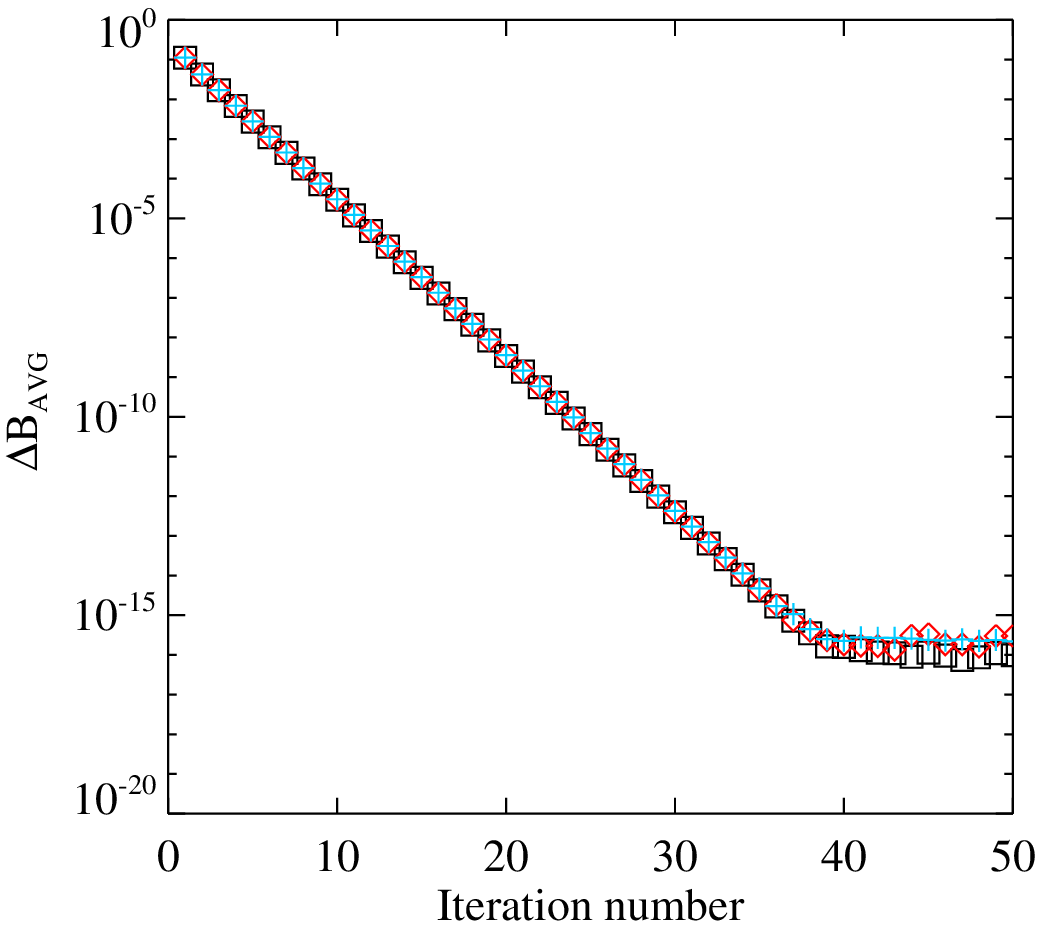}}
\caption{The mean absolute change in the field $\Delta B_{\rm avg}$ 
as a function of iteration number for the application of the Grad-Rubin
method to the test cases. The left panel shows $\Delta B_{\rm avg}$
for the three tests with periodic side boundary conditions, and  
the right panel shows $\Delta B_{\rm avg}$ for the three tests 
with closed boundary conditions. The different symbols 
represent $\Delta B_{\rm avg}$ for different grid 
sizes $N$. In both panels the $N=65$ case is shown with squares, the 
$N=101$ case is shown with diamonds, and the $N=151$ case is 
shown with plus signs, and the scale on the $y$ axis is logarithmic.} 
\label{f2}
\end{figure}

The convergence of the Grad-Rubin 
iteration is measured by the absolute change in the 
field at each iteration, {\it i.e.}
\BE
  \Delta B_{\rm avg} = \langle | \vec B^{(k)} - \vec B^{(k-1)}| \rangle 
  \label{abchg}
\EE
where $\langle  \rangle$ denotes the average over all points in the 
computational volume. The left panel of Figure \ref{f2} 
shows $\Delta B_{\rm avg}$ over 
30 Grad-Rubin iterations for the three different grid sizes. 
The squares show the case with $N=65$, the diamonds show
the case with $N=101$, and the plus signs show the case with 
$N=151$. In all three cases the Grad-Rubin iteration
converges: $\Delta B_{\rm avg}$ decreases exponentially for 
 $\approx 15$ iterations before becoming approximately constant. The rate of 
convergence does not appear to depend strongly on grid size.

\begin{figure}[!h]
 \centerline{\includegraphics[scale =1.]{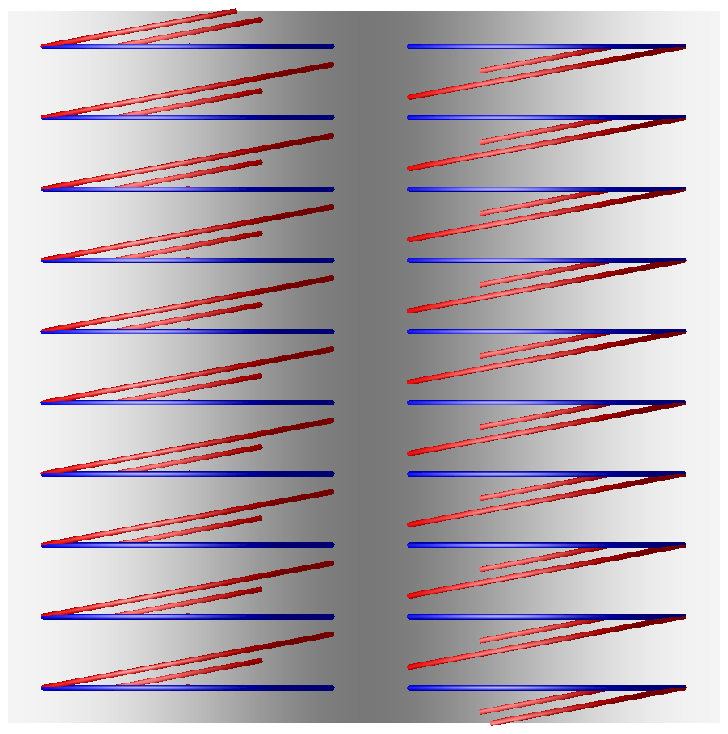}\hspace{1cm}\includegraphics[scale =1.]{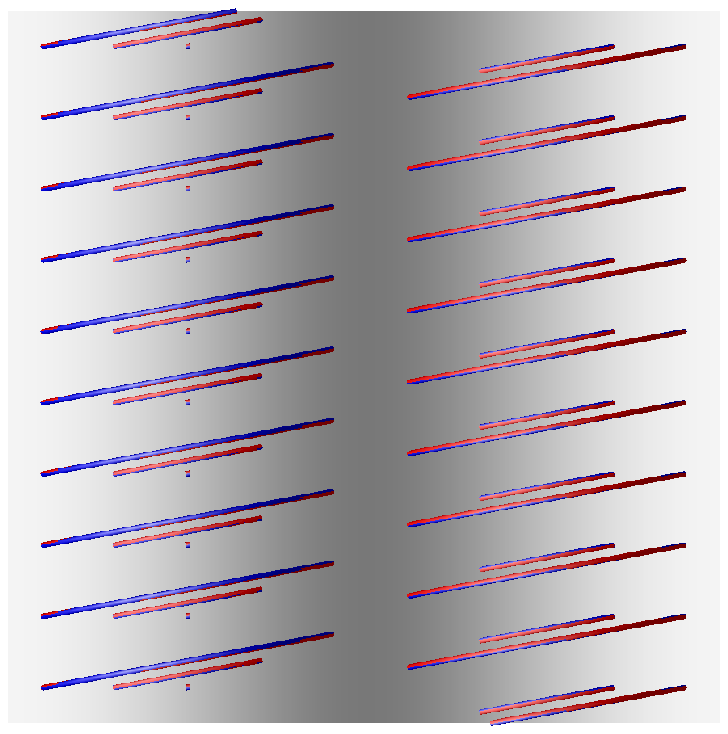}}
\caption{Comparisons between the field lines for the analytic solution, 
the Grad-Rubin solution and the initial potential field for the test case 
calculation in Section \ref{TC1}. The left panel 
shows the analytic solution 
(red field lines) and the potential field (blue field lines). 
The right panel shows the analytic solution (red field lines) 
and the numerical solution after 30 Grad-Rubin iterations 
(blue field lines). The solutions are viewed looking down 
from the top of the computational domain. 
The $z=0$ plane is shaded to show $B_{\rm obs}$ (regions with 
$B_{\rm obs}<0$ are dark and those with $B_{\rm obs}>0$ are light). 
The solution after 30 Grad-Rubin iterations closely matches the analytic solution. } 
\label{f3}
\end{figure}

The numerical solution after 30 iterations is compared to the 
analytic solution. This is done qualitatively by
comparing the field lines of the analytic solution, the initial potential
field, and the numerical solution. The left panel of Figure \ref{f3} shows
the field lines of the analytic solution (red field lines), and the field lines of the 
potential field (blue field lines) viewed looking down on the computational domain. 
There is a significant difference between the two sets of field lines. 
The right panel of Figure \ref{f3} shows the field lines of the 
analytic solution (red field lines), and the field lines of the numerical 
solution after 30 Grad-Rubin iterations (blue field lines) from the same 
viewpoint. The two sets of field lines closely coincide, 
confirming that the Grad-Rubin iteration has converged to 
the analytic solution. 

In addition to this qualitative comparisons, we compare the analytic
and numerical solutions quantitatively. Several metrics have been
developed for this purpose in the context of nonlinear force-free modeling
\citep{2006SoPh..235..161S}. The various metrics show similar results, so
for brevity we present only the mean vector error

\BE
  E_{\rm m} = \left \langle \frac{|\vec B-\vec b|}{|\vec B|} \right \rangle,
  \label{mve}
\EE
and the metric 
\BE
E_{\rm CS} = 1-\left \langle \frac{\vec B \cdot \vec b}{|\vec B||\vec b|} \right \rangle,
\label{cs} 
\EE
which is based on the Cauchy-Schwartz inequality\footnote{These metrics 
are chosen because they show the largest discrepancy between the numerical 
and the analytic fields.}. In both these definitions the analytic solution 
is $\vec B$, the numerical solution is $\vec b$, and the average is over all grid points in the 
domain. For two exactly matching fields $E_{\rm m}=0$ and $E_{\rm CS}=0$.
The mean vector error is a measure of the difference between
the magnitude and direction of $\vec B$ and $\vec b$ at each 
grid point, while $E_{\rm CS}$ is only sensitive to the differences
in the direction of the fields. 

We also compute 
\BE
E_{\rm div} = \langle | \nabla \cdot \vec b | \rangle, 
\label{div_av}
\EE
which is a measure of the divergence of the numerical solution $\vec b$.
A second-order finite difference approximation is used to compute
the divergence at each grid point \citep{1992nrfa.book.....P}. In principle 
$E_{\rm div}$ should be zero, but this is not achieved in practice
due to the finite numerical accuracy of the solution, and the truncation
error introduced by the numerical approximation to differentiation. 
The truncation error in the derivative has a scaling $\sim 1/N^2$ 
\citep{1992nrfa.book.....P}. 

\begin{figure}[!h]
\centerline{\includegraphics[scale =0.8]{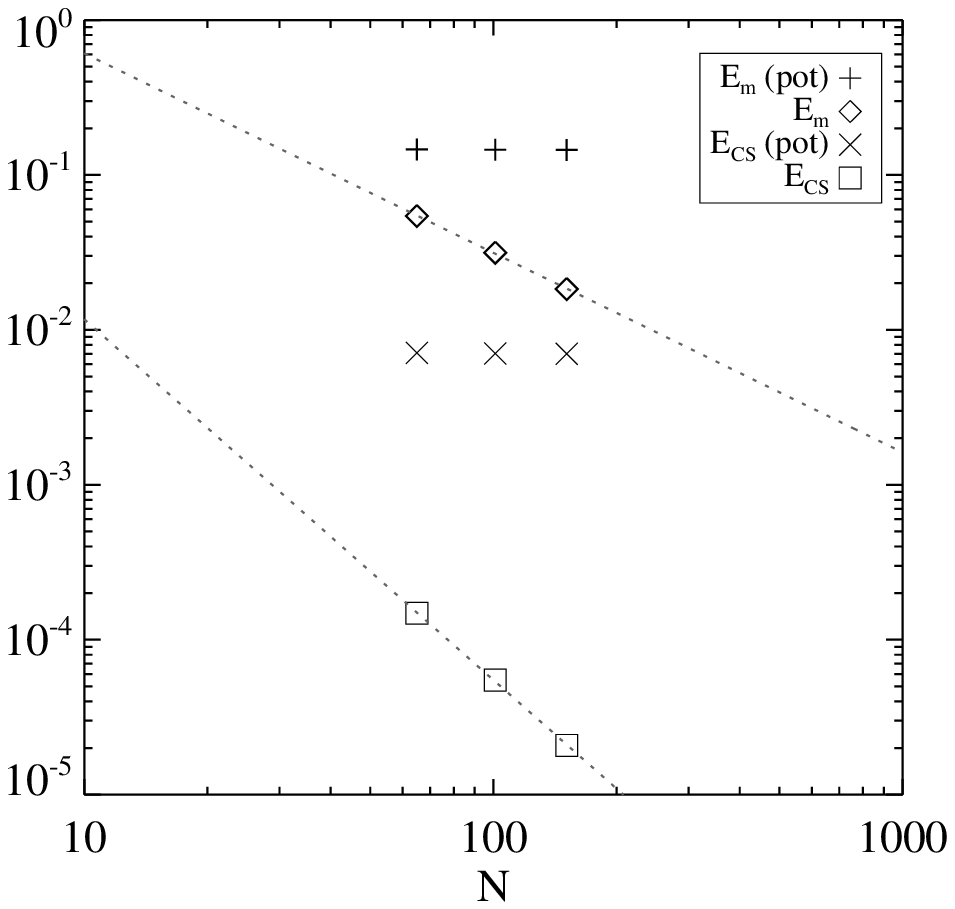} \includegraphics[scale =0.8]{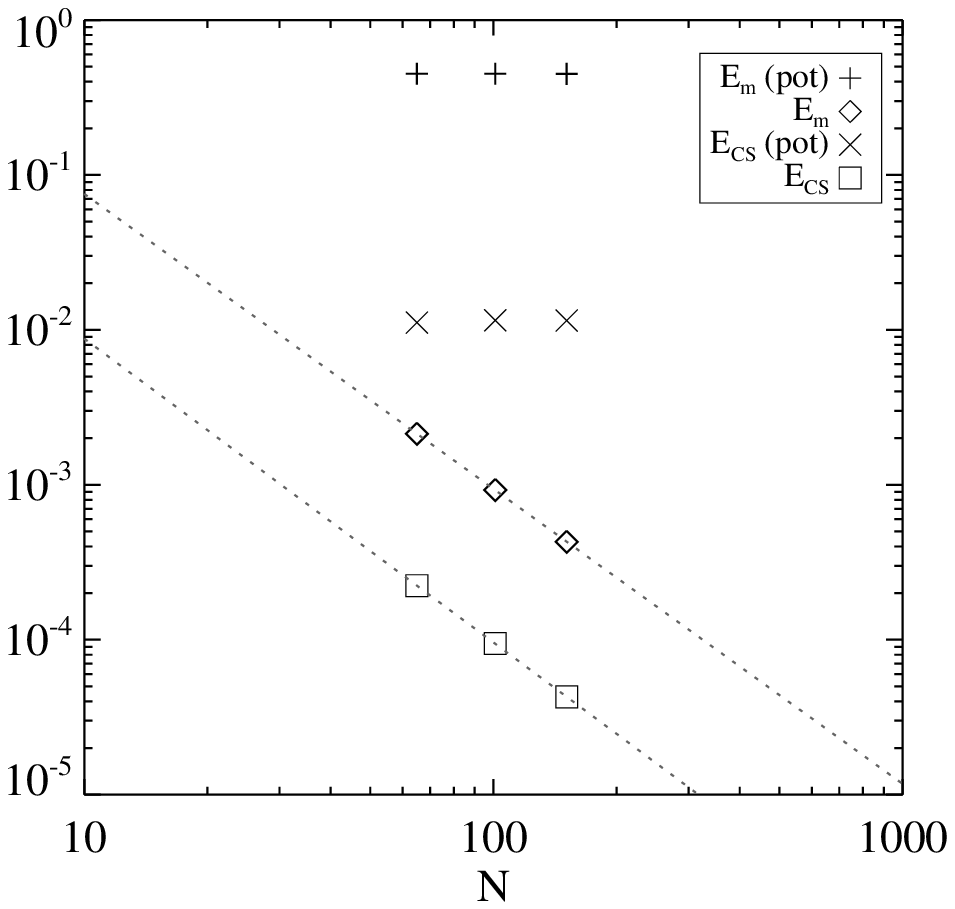}}
\caption{Error metrics $E_{\rm m}$ (see Equation \eq{mve}) and $E_{\rm CS}$
(see Equation \eq{cs}) as a function of grid size $N$ for the 
initial potential field (plus signs and crosses), and for the numerical solution 
(diamonds and squares). 
The left panel shows the results for the test cases
with periodic side boundary conditions after 30 Grad-Rubin iterations, 
and the right panel shows the results for the test cases with closed 
side boundary conditions after 50 Grad-Rubin iterations. The Grad-Rubin 
solutions show power-law scaling $E_{\rm m} \sim N^{\gamma}$ and 
$E_{\rm CS} \sim N^{\gamma}$, and the dashed lines show power-law
fits to the data. The power-law indices
$\gamma$ for each fit are summarized in Table \ref{t1}.}
\label{f4}
\end{figure}

The left panel of Figure \ref{f4} shows $E_{\rm m}$ and $E_{\rm CS}$ 
for the initial potential field (plus signs and crosses), and for 
the numerical solution after 30 Grad-Rubin iterations (diamonds
and squares), as functions of grid size $N$. The error associated 
with the potential field is independent of $N$. For the Grad-Rubin solution both $E_{\rm m}$ 
and $E_{\rm CS}$ decrease approximately as power laws, {\it i.e.} show 
scalings $\sim N^{\gamma}$, where we find (based on least squares
fits) $\gamma = -1.3$ for $E_{\rm m}$ and $\gamma = -2.3$ for $E_{\rm CS}$.
The power-law fits are shown in Figure \ref{f4} by the dashed lines. 
The scaling for $E_{\rm m}$ is close to the expected $\sim 1/N$ scaling discussed
in Section \ref{steps_23}, while the scaling for $E_{\rm CS}$ is closer to $\sim 1/N^2$.  

\begin{figure}[!h]
\centerline{\includegraphics[scale =0.8]{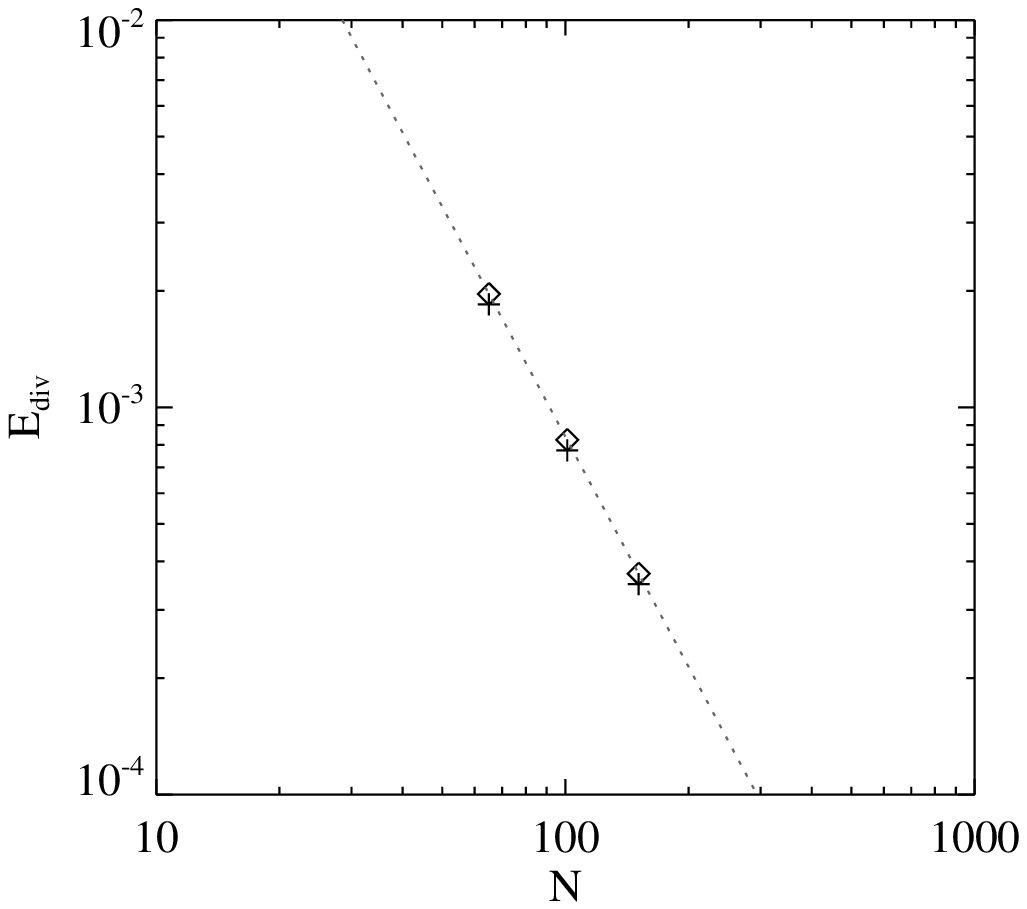} \includegraphics[scale =0.8]{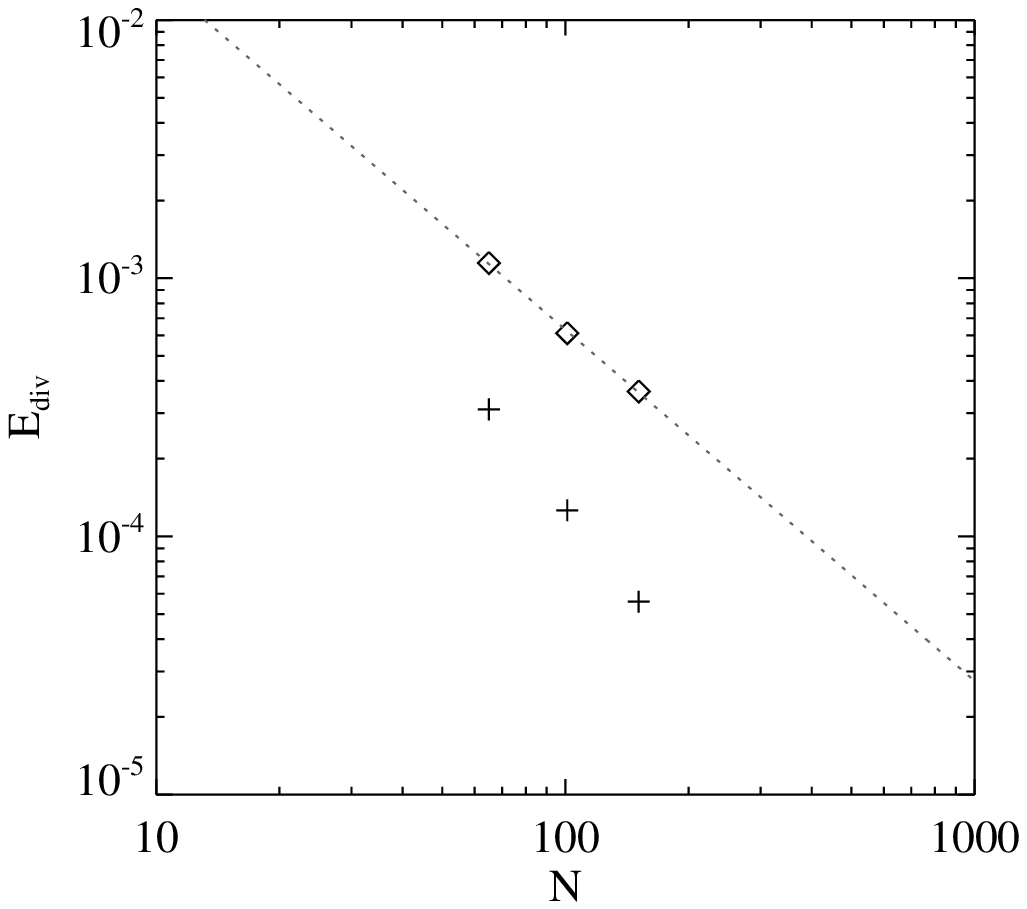}}
\caption{The absolute value of the divergence of the 
numerical solution and initial potential fields, averaged over the 
computational domain $E_{\rm div}$ (see Equation \eq{div_av}) as a function of grid size $N$. 
The left panel shows $E_{\rm div}$ for the analytic solution (plus signs) 
and the numerical solution after 30 Grad-Rubin iterations for the test case with periodic 
boundary conditions. The right panel shows $E_{\rm div}$
for the analytic solution (plus signs) and the numerical solution
after 50 Grad-Rubin iterations for the test case with closed
boundary conditions. The dashed lines are power-law fits ($\sim N^{\gamma}$)
to $E_{\rm div}$ for the numerical solutions. The power-law indices
are $\gamma = -2.0$ for the data in the left panel, and
$\gamma = -1.4$ for the data in the right panel. The power-law index 
for $E_{\rm div}$ for the analytic solutions in both panels is $\gamma=-2.0$.
The fits are not shown. }
\label{f5}
\end{figure}

The left panel of Figure \ref{f5} shows $E_{\rm div}$ as a function
of $N$ for the numerical solution after 30 Grad-Rubin iterations
(diamonds) and the analytic solution (plus signs). The the two 
data sets overlap indicating that the divergence of the numerical 
solution is close to or smaller than the truncation error in the numerical 
derivative. In common with $E_{\rm m}$ and $E_{\rm CS}$, the metric $E_{\rm div}$ 
for the Grad-Rubin solution has power-law scaling $\sim N^{\gamma}$. 
We estimate $\gamma=-2.0$ based on a least squares fit to the data for the 
Grad-Rubin solution (the power-law fit is shown as a dashed line in 
the figure). 

\begin{figure}[!h]
\centerline{\includegraphics[scale =1.]{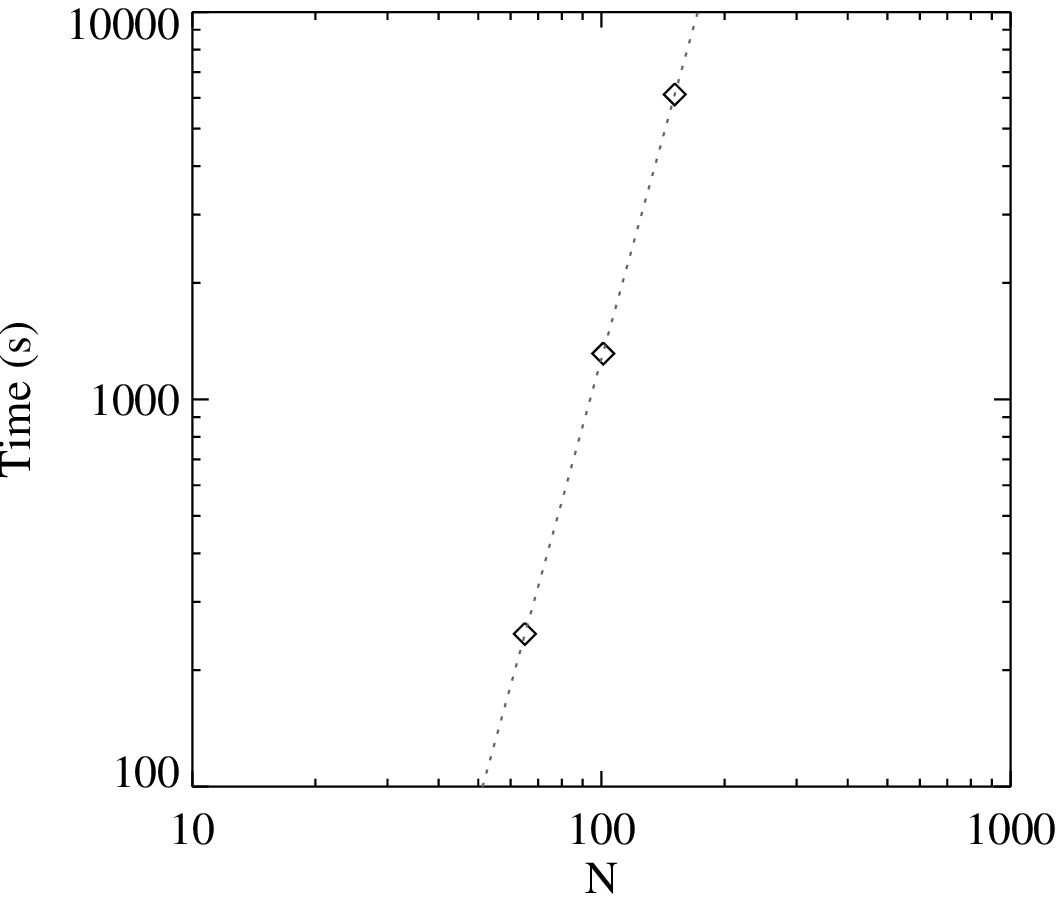}}
\caption{Execution time for 30 Grad-Rubin iterations
as a function of grid size $N$ for the first test case (Section \ref{TC1}). 
The execution time has power-law scaling with a  
power-law index $\gamma = 3.8$ based on a least-squares 
fit to the data (the dashed line).}
\label{f6}
\end{figure}

We also measure the run time of the code for different grid sizes using the
CPU clock (the tests are performed on an eight core CPU). 
Figure \ref{f6} shows the execution time for 30 Grad-Rubin iterations
as a function of $N$. The results appear to follow a power law 
$\sim N^{\gamma}$, and we estimate $\gamma = 3.8$ 
based on a least squares fit to the three data points (the fit is shown
by the dashed line). This scaling is close to the $\sim N^4$ scaling expected for 
the field line tracing discussed in Section \ref{steps_23}. 
This scaling implies that the field line tracing is the computationally slowest 
step in the calculation.

\subsection{Test Case with Closed Side Boundaries}
\label{TC2}

For the second test we use the parameters
$k=\pi/L$, $\lambda=0.9\pi/L$, and $a_0=1$. The grid
sizes are the same as in Section \ref{TC1}, and we 
apply 50 Grad-Rubin iterations starting from a potential field.

The right panel of Figure \ref{f2} shows $\Delta B_{\rm avg}$
over 50 Grad-Rubin iterations for the three grid sizes used. 
For all three cases $\Delta B_{\rm avg}$ decreases exponentially
for roughly 40 iterations before becoming approximately constant. The 
rate of convergence does not appear to depend strongly on the grid
size. 

\begin{figure}[!h]
 \centerline{\includegraphics[scale =0.50]{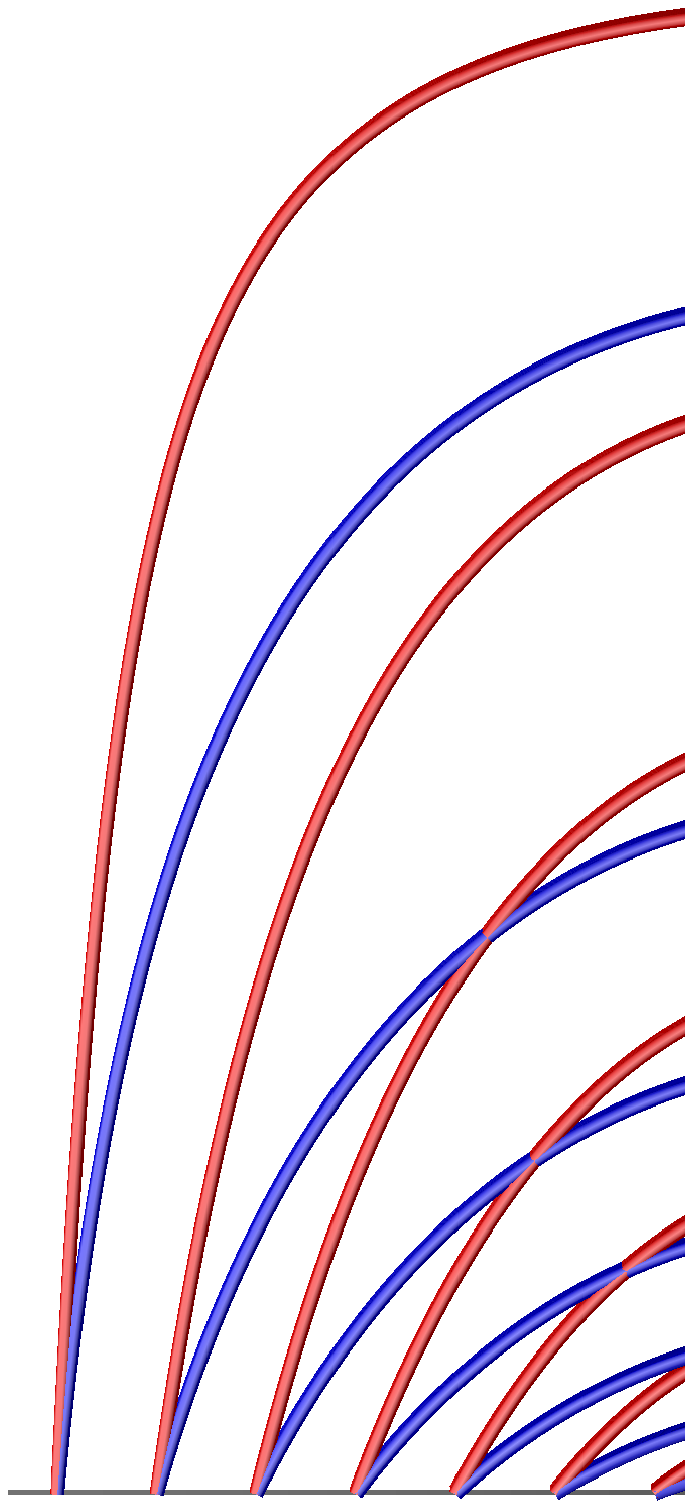}\hspace{1cm} \includegraphics[scale =1.]{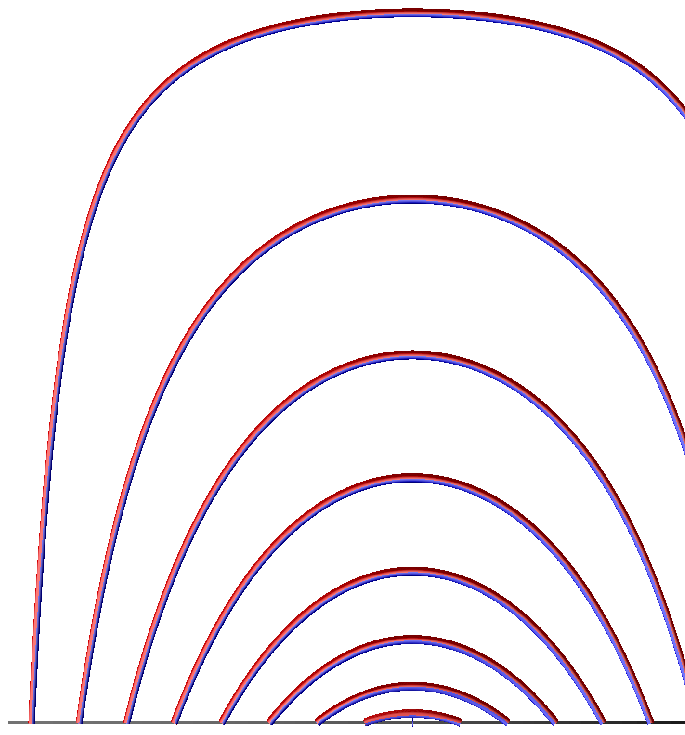}}
 \caption{Comparison between the field lines for the analytic solution, 
the Grad-Rubin solution and the initial potential field for the test case in
Section \ref{TC2}. The left panel shows the analytic 
solution (red field lines) and the potential field (blue field lines). 
The right panel shows the analytic solution (red field lines)
 and the numerical solution after 50 Grad-Rubin iterations 
(blue field lines).  The view is along the $y$ axis in both panels. 
The Grad-Rubin solution closely matches the analytic solution. } 
\label{f7}
\end{figure}

Figure \ref{f7} shows the field lines of the 
analytic solution, the potential field, and the Grad-Rubin solution.
The view in the two panels in the figure is along the $y$ axis. The left panel
shows the field lines of the initial potential field (blue field lines) and of the
analytic solution (red field lines). The right panel shows the field lines of 
the Grad-Rubin solution after 50 Grad-Rubin iterations (blue field lines) and of
the analytic solution (red field lines). The Grad-Rubin solution closely 
matches the analytic solution.  

The right panel of Figure \ref{f4} shows the error metrics $E_{\rm m}$ and $E_{\rm CS}$ 
for the initial potential field (plus signs and crosses), and for 
the numerical solution after 50 Grad-Rubin iterations (diamonds
and squares). The error associated with the potential field is 
independent of $N$. For the Grad-Rubin solution both $E_{\rm m}$ 
and $E_{\rm CS}$ decrease as power-laws, with indices 
$\gamma = -1.9$ for $E_{\rm m}$ and $\gamma = -2.0$ for $E_{\rm CS}$ (based on least-squares
fits, shown by the dashed lines). Both error metrics are found to scale 
approximately as $1/N^2$. The expected scaling, discussed in Section \ref{steps_23}, 
is $\sim 1/N$. The improvement in the observed scaling may be due 
to the simplicity of the solution (for this solution $B_y=0$) and 
is unlikely to occur more generally. 

The right panel of Figure \ref{f5} shows the error metric $E_{\rm div}$ as a function
of $N$ for the numerical solution after 50 Grad-Rubin iterations
(diamonds) and for the analytic solution (plus signs). In this case 
$E_{\rm div}$ for the Grad-Rubin solution has power-law with index
$\gamma=-1.0$ (based on a least squares fit to the data, shown as a 
dashed line in the figure). The analytic solution has a power-law 
scaling in $E_{\rm div}$ with $\gamma=-2.0$ (the fit is not
shown). There is a substantial difference between $E_{\rm div}$ for 
the analytic and numerical solution indicating that $E_{\rm div}$
is providing a measure of the residual divergence of the numerical solution and
is not due to the truncation error incurred by the numerical approximation
to differentiation.

%
%
\section{Discussion and Conclusion}
\label{discussion}

We present an implementation of the Grad-Rubin method for 
solving the magnetostatic equations in a finite domain. 
The code allows two possible choices of boundary conditions 
on the side boundaries of the domain: either $\vec B$ is 
periodic on the side boundaries, or the normal component of 
$\vec B$ is zero on the side boundaries. We refer to these 
choices as periodic boundary conditions and closed
boundary conditions. In both cases the normal component of $\vec B$ is
zero on the top boundary of the domain.

The code is tested in application to a simple analytic solution, for two 
choices of parameters for the solution, which illustrate the 
periodic and closed boundary conditions respectively. In both cases the 
code accurately reconstructs the test solution. Several runs are performed 
for each test case with varying grid sizes, to demonstrate the scaling
of the method with the size of the problem. The test case is 
highly idealized. The lower boundary conditions are exactly 
consistent with the magnetostatic model because they are
derived from an exact solution, and the top and side boundary conditions
exactly match the assumptions adopted for the numerical method.
The idealization allows a rigorous test of the correctness of the implementation.
 
For both test cases the Grad-Rubin method converges.
The convergence is measured using the average absolute 
change $\Delta B_{\rm avg}$ in the field (see Equation \eq{abchg}). 
The test case with periodic boundary conditions converges faster than
the test case with closed boundary conditions, measured in the number of 
iterations. This is likely due to the second test
case being significantly more non-potential than the first, rather
than being due to the different boundary conditions.

The code is found to accurately reproduce
the analytic test cases. This is confirmed by a visual 
comparison of the field lines of the analytic solution and
the reconstructed solution. For both choices of 
parameters/boundary conditions the numerical solution succeeds, 
based on this test. We also measure the success using the mean 
vector error $E_{\rm m}$ which is defined by Equation \eq{mve}, and the Cauchy-Schwartz
inequality based metric $E_{\rm CS}$, which is defined by 
Equation \eq{cs}. We also compute a measure of the residual divergence
$E_{\rm div}$ which is defined by Equation \eq{div_av}. In the absence
of numerical error we would expect $E_{\rm m}=E_{\rm CS}=E_{\rm div}=0$. 
For both test cases $E_{\rm m}$, $E_{\rm CS}$, and $E_{\rm div}$ 
decrease as $N$ increases, with a power-law scaling in $N$, and we estimate
the power-law index in each case from a least-squares fit to the
data. The power-law indices are summarized in Table \ref{t1}. We
obtain different scalings for the different metrics in each case.
This may be attributed to the difference between the two test cases, 
and the different metrics. The slowest scaling achieved is $\sim 1/N$ 
in each case, which is consistent with other Grad-Rubin implementations 
\citep{2006SoPh..238...29W}, and with the estimate made in Section \ref{steps_23}.
The fastest scaling achieved is $\sim 1/N^2$, which may be attributed
to the simple form of the analytic solutions and is unlikely to be 
achieved for more general solutions. In general we expect a scaling 
$\sim 1/N$. 

\begin{table}[!h]
\caption{Power law indices $\gamma$ for the scaling of the error 
metrics $E_{\rm m}$ , $E_{\rm CS}$, and $E_{\rm div}$ with grid size.}
\begin{tabular}{l l  l  l }
\hline
\hline
Test case & $E_{\rm m}$  & $E_{\rm CS}$  & $E_{\rm div}$  \\
\hline
Periodic & -1.3  & -2.3 & -2.0    \\
Closed & -1.9  & -2.0 & -1.4     \\

\end{tabular}
\label{t1}
\end{table}

The execution time of the code is found to scale as $\sim N^4$, which
is comparable to the fastest force-free Grad-Rubin 
methods ({\it e.g.} \citealt{2006SoPh..238...29W,2007SoPh..245..251W}).
However, the magnetostatic implementation is significantly slower
in absolute terms than the force-free methods. The slowest step 
in both cases (force-free and magnetostatic) is the field line tracing
used to update quantities in the volume at each iteration,
and the magnetostatic case has two field line tracing steps 
(for $\sigma^{(k)}$ and $p^{(k)}$) compared to 
one (for $\alpha^{(k)}$) in the force-free case.

This paper demonstrates the new method and code in application
to a simple test case. The long-term goal of our work is to 
develop a method and code for reconstructing coronal magnetic
fields for real solar data. Several 
obstacles remain to be overcome before this can be achieved. 
One problem is observational: the photospheric pressure profile 
is not currently available for real solar cases. However, progress 
might be made by assuming a simple model for the pressure, {\it e.g.} 
a constant $\beta$ model, with $p \propto |\vec B|^2$. 
Our model is also overly simplified in that we neglect gravity, and
include only a pressure force. We are presently considering ways
to include gravity also. Finally, it remains to be seen
how well the techniques being developed cope with noisy data, and with 
data which are not strictly consistent with the magnetostatic model.
Despite these qualifications, the method outlined here, and its successful
application to analytic test cases, represents an important first step towards a Grad-Rubin 
method for magnetostatic reconstructions 
of the coronal magnetic field.

%
%
\section*{Appendix A}

A potential field $\vec B_0$ satisfying 
\BE
  \nabla \times \vec B_0 = 0
\EE
and the boundary conditions at $z=0$ 
\BE
  \vec B_0 \cdot \ut{z} = B_{\rm obs}
\EE
is used to initiate the Grad-Rubin iteration. The appropriate 
field with periodic side boundary conditions is 
given in Equation \eq{b01}-\eq{b03} in the text. For the
choice of closed side boundaries the appropriate choice is 
\begin{eqnarray}
B_{0x}(x,y,z) &=& \sum_{m=0}^{\infty}\sum_{n=0}^{\infty} c_{mn} k_m
                  \cosh(k[z-L])\sin(k_m x)\cos(k_n y),\label{apAb1}\\
B_{0y}(x,y,z) &=& \sum_{m=0}^{\infty} \sum_{n=0}^{\infty} c_{mn} 
                  k_n \cosh(k[z-L])\cos(k_m x)\sin(k_n y),\label{apAb2}
\end{eqnarray}
and
\BE
B_{0z}(x,y,z) = -\sum_{m=0}^{\infty} \sum_{n=0}^{\infty} c_{mn} k
                   \sinh(k[z-L])\cos(k_m x)\cos(k_n y)\label{apAb3}, 
\EE
where $k_m = \pi m/L$, $k_n = \pi n/L$, $k^2 = k_m^2+k_n^2$, and 
where the coefficients $c_{mn}$ are given by 
\BE
 c_{mn} = \frac{4}{L^2} \int_0^{L} \int_0^{L} dxdy 
 B_{\rm obs}(x,y) \cos(k_mx)\cos(k_ny)/\sinh(kL)\label{apAb4}.
\EE
Equations \eq{apAb1}-\eq{apAb4} may be evaluated on a grid with
$N^3$ points in $\sim N^3 \log(N)$ operations using fast sine and 
cosine transforms \citep{1996tah..book.....P}.

%
%
\section*{Appendix B}

This appendix presents the solution for the current-carrying 
component of the test field, $\vec B_c$, with closed top and side boundary 
conditions. The magnetic field may be calculated from a vector potential $\vec A$
in the Coulomb gauge ($\nabla \cdot \vec A=0$ ) using 
$\nabla \times \vec A = \vec B_c$. The vector potential satisfies the
vector Poisson equation \citep{1998clel.book.....J}: 
\BE
\nabla ^2 \vec A = -\mu_0 \vec J.
\label{poisson_eq11}
\EE
The boundary conditions for $\vec B_c$ on the six plane boundaries are 
\BE
  \vec B_c \cdot \ut{n} = 0,
\EE
where $\ut{n}$ is the unit vector normal to the boundary. The corresponding
boundary conditions for the vector potential in the Coulomb 
gauge are \citep{1999A&A...350.1051A}:  
\BE
  \partial_n \vec A =0,
  \label{apB1}
\EE 
and
\BE
  \vec A_t =0,
  \label{apB2}
\EE
where $\vec A_t$ denotes the component of $\vec A$ 
transverse to the boundary, and $\partial_n \vec A$ denotes the 
normal derivative.  

The vector potential $\vec A$ satisfying the boundary conditions Equations
\eq{apB1}-\eq{apB2} can be written as a Fourier series: 

\begin{eqnarray}
A_x &=& \tsum a^{(1)}_{mnp} 
        \cos(k_mx)\sin(k_ny)\sin(k_pz), \label{apB3} \\
A_y &=& \tsum a^{(2)}_{mnp} 
       \sin(k_mx)\cos(k_ny)\sin(k_pz), \label{apB4}
\end{eqnarray}
and
\BE
A_z = \tsum a^{(3)}_{mnp} 
       \sin(k_mx)\sin(k_ny)\cos(k_pz), \label{apB5}
\EE
where $k_m = \pi m/L$, $k_n = \pi n/L$, and $k_p = \pi m/L$.
In the following we solve the Poisson equation 
to find an expression for $a_{mnp}$. The process is demonstrated for
the $A_x$ component, but the approach is similar for the other 
components. 

Substituting Equation \eq{apB3} into the Poisson equation (Equation \eq{poisson_eq11}) gives 
\BE
A_x =\tsum a^{(1)}_{mnp} k^2
     \cos(k_mx)\sin(k_ny)\sin(k_pz) = \mu_0 J_x(x,y,z),
\EE
where $k^2 = k_m^2+k_n^2+k_p^2$.

Applying the standard orthogonality relations (where $\delta_{mn}$ is 
the Kronecker delta): 
\begin{eqnarray}
\int_{0}^{L} \sin(\pi m s/L)\sin(\pi n s/L)ds &=& \frac{L}{2} \delta_{mn},\\
\int_{0}^{L} \cos(\pi m s/L)\cos(\pi n s/L)ds &=& \frac{L}{2} \delta_{mn},
\end{eqnarray}
and
\BE
  \int_{0}^{L} \sin(\pi m s/L)\cos(\pi n s/L)ds = 0
\EE
yields
\BE
a^{(1)}_{mnp} = \frac{8\mu_0}{L^3}
                \int_{0}^{L}\int_{0}^{L}\int_{0}^{L}
                J_x(x,y,z) \cos(k_mx)\sin(k_ny)\sin(k_pz)dxdydz.
\EE
Similar expressions apply for the other two coefficients: 
\BE
  a^{(2)}_{mnp}  = \frac{8\mu_0}{L^3} 
                   \int_0^{L}\int_0^{L}\int_0^{L} 
                    J_y(x,y,z) \sin(k_m x)\cos(k_n y)\sin(k_p z)dxdydz, \\
\EE
and
\BE
  a^{(3)}_{mnp}  = \frac{8\mu_0}{L^3} 
                  \int_0^{L}\int_0^{L}\int_0^{L} 
                  J_z(x,y,z) \sin(k_m x)\sin(k_n y)\cos(k_p z)dxdydz.
\EE
The magnetic field is obtained by evaluating 
$\vec B_c = \nabla \times \vec A $. The components 
of $\vec B_c$ are 

\begin{eqnarray}
B_{cx}(x,y,z) &=& \tsum \left [k_na^{(3)}_{mnp}-k_pa^{(2)}_{mnp} \right ]
                  \sin(k_m x)\cos(k_n y)\cos(k_p z)/k^2, \label{apB6}\\
B_{cy}(x,y,z) &=& \tsum \left [k_pa^{(1)}_{mnp}-k_ma^{(3)}_{mnp} \right ]
                  \cos(k_m x)\sin(k_n y)\cos(k_p z)/k^2,\label{apB7} \\ 
\end{eqnarray}
and
\BE
B_{cz}(x,y,z) = \tsum \left [k_m a^{(2)}_{mnp}-k_n a^{(1)}_{mnp}\right ]
                  \cos(k_m x)\cos(k_n y)\sin(k_p z) /k^2 \label{apB8}.  
\EE
The solution given by Equations \eq{apB6}-\eq{apB8} may be 
computed on a grid with $N^3$ points in $\sim N^3\log(N)$ operations using a 
combination of fast sine and cosine transforms.




\begin{acks}
S. A. Gilchrist acknowledges the support of an Australian 
Postgraduate Research Award. 
\end{acks}


\end{article} 
\end{document}